\documentclass[referee]{raa}
\usepackage{graphicx,times}
\usepackage[section]{placeins}

\usepackage{epsfig}
\usepackage{makeidx}
\usepackage{ifthen}
\usepackage{verbatim}
\usepackage{listings}
\usepackage{float}

\usepackage{hyperref}

\usepackage{natbib}
\usepackage{amssymb,amsmath}
\usepackage{rotating}
\bibpunct{(}{)}{;}{a}{}{,}

\newcommand{\kms}{kms$^{-1}$}
\newcommand{\chou}{C07}
\newcommand{\belokurov}{B14}
\newcommand{\lm}{LM10}
\newcommand{\li}{Paper I}

\newcommand{\ngc}{\emph{NGC}}
\newcommand{\ww}{$(W_1-W_2)_0$}
\newcommand{\structA}{\emph{Vel-3+83}}
\newcommand{\structB}{\emph{Vel+162+26}}
\newcommand{\Lsgr}{$\rm\Lambda$}
\newcommand{\Bsgr}{$\rm B$}
\newcommand{\vgsr}{$V_{gsr}$}
\bibpunct{(}{)}{;}{a}{}{,} 
\hypersetup{
pdftitle={Substructure nearby the Sagittarius stream},
pdfauthor={Jing Li},
colorlinks=true,
linkcolor=blue,
citecolor=blue,
urlcolor=blue
}

\begin{document}

\title{New tidal debris nearby the Sagittarius leading tail from the \textbf{LAMOST DR2 M giant stars}
$^*$
\footnotetext{\small $*$ Supported by the Strategic Priority Research Program ``The Emergence of Cosmological Structures" of the Chinese Academy of Sciences No. XDB09000000 and the National Key Basic Research Program of China 2014CB845700.}
}

 \volnopage{ {\bf 2016} Vol.\ {\bf X} No. {\bf XX}, 000--000}
   \setcounter{page}{1}
   
\author{
Jing Li\inst{1,2}, Chao Liu\inst{3}, Jeffrey L. Carlin\inst{4}, Jing Zhong\inst{1}, Jinliang Hou\inst{1}, Licai Deng\inst{3}, Heidi Jo Newberg\inst{4}, Yong Zhang\inst{5}, Yonghui Hou\inst{5}, Yuefei Wang\inst{5}}

\institute{Key Laboratory of Galaxies and Cosmology,
Shanghai Astronomical Observatory, Chinese Academy of Sciences,
80 Nandan Road, Shanghai,200030, China;{\it lijing@shao.ac.cn}\\
\and
University of Chinese Academy of Sciences, 19A Yuquan Road, Beijing 100049, China;
\and 
Key Laboratory of Optical Astronomy, National Astronomical Observatories, Chinese Academy of Sciences, Beijing 100012;{\it liuchao@nao.cas.cn}
\and
Dept. of Physics, Applied Physics and Astronomy, Rensselaer Polytechnic Institute Troy, NY 12180;
\and
Nanjing Institute of Astronomical Optics \& Technology, National Astronomical Observatories, Chinese Academy of Sciences, Nanjing; 210042
}

\abstract{
We report two new tidal debris nearby the Sagittarius (Sgr) tidal stream in the north Galactic cap identified from the M giant stars in LAMOST DR2 data. The M giant stars with sky area of $210^\circ<$\Lsgr$<290^\circ$, distance of 10--20\,kpc, and [Fe/H]$<-0.75$ show clear bimodality in velocity distribution. We denote the two peaks as \structA\ for the one within mean velocity of -3\,\kms\ with respect to that of the well observed Sgr leading tail at the same \Lsgr\ and \structB\ for the other one with mean velocity of 162\,\kms\ with respect to the Sgr leading tail. Although the projected \Lsgr--\vgsr\ relation of \structA\ is very similar to the Sgr leading tail, the opposite trend in \Lsgr--distance relation against the Sgr leading tail suggests \structA\ has a different 3D direction of motion with any branch of the simulated Sgr tidal stream from Law \& Majewski. Therefore, we propose it to be a new tidal debris not related to the Sgr stream. Similarly, the another substructure \structB, which is the same one as the \ngc\ group discovered by Chou et al.,  also moves toward a different direction with the Sgr stream, implying that it may have different origin with the Sgr tidal stream.
}

  \authorrunning{J. Li et al. }            %author_head in even pages
   \titlerunning{New tidal debris nearby the Sagittarius leading tail }  % title_head in odd pages
   \maketitle

\section{Introduction}

The $\Lambda CDM$ cosmology predicts that the halo of  Milky Way like galaxies should contain hundreds of sub-haloes,  in which dwarf galaxies may be embedded. Many dwarf galaxies and tidal substructures in the Galactic halo have been discovered in the last decade~\citep[][etc.]{nyetal02,rochapinto2004,Bel06,2009ApJ...700L..61N,bonaca2012}. The most prominent one is the Sagittarius (Sgr) stream, which is believed to be a tidal debris of the disrupting Sgr dwarf galaxy. After it was firstly discovered by \citet{Mateo1996}, the stream has been mapped over $2\pi$ radians on the sky by the 2MASS \citep{Majewski2003} and SDSS \citep{Bel06,Koposov2012} surveys. The stream, which is composed of the leading and the trailing tails, wraps at least once around the Galaxy. Some modeling works even claimed that the stream should wrap more than once \citep{Penarrubia2010,2010ApJ...714..229L}. One prominent feature of the Sgr stream is the bifurcations in both north and south hemisphere.

Other than the large scale and prominent substructures, there are also some local and less prominent substructures in the region of the Sgr stream. Some of them may be related to the Sgr stream, some may be new tidal debris with different origins. \citet[][hereafter \chou]{Chou2007} found two substructures in the region of the Sgr leading stream using M giant stars. One is likely the part of the Sgr leading tail with similar metallicity and velocity. The authors further separated the tracer stars as the \emph{best} and the \emph{less certain} subsamples. The former are located between 10 and 20\,kpc, while the latter is located within 5\,kpc in distance. They thought most of the best and less subsample should belong to the Sgr stream because this region is far way from the Galactic disk and no evidence for the existence of other tidal substructure is found from the star count of the M giant stars. The other substructure, which is denoted as the north Galactic cap (\ngc) group by the authors, shows opposite velocity to the Sgr leading tail and is proposed to be the Sgr trailing tail overlapped with the leading tail in the north.

\citet{Pila2014} photometrically observed three fields around the Sgr leading tail and found that the distance of the Sgr leading tail well matches the nearer branch in the model from \citet{Penarrubia2010}. They claimed that this belongs to a new wrap of the Sgr tidal stream together with another branch in the trailing region claimed in the same paper. However, the distance from \citet{Pila2014} is not consistent with from the north trailing tail in the simulation from~\citet[][hereafter \lm]{2010ApJ...714..229L}, requiring more kinematical data to confirm their conclusion.

It is noted that \citet{heidi07} explored an overdensity at (\Lsgr, $g_0$) $=$ ($240^\circ$, $16.7$\,mag), whose origin is unclear, from the blue horizontal branch stars in the same region of the Sgr leading tail. $g_0$ is the g-band apparent magnitude with reddening corrected. Addition, they also found a moving group near S297+63-20.5 \footnote{The "S" signifies stream (or substructure), followed by the Galactic longitude, then Galactic latitude, and the apparent magnitude of the reddening corrected $g_0$ turnoff.\cite{heidi07} } with $Vgsr=-76 \pm 10$\,\kms\ and distance of $15.8$ kpc to the Sun. Since the location is quite close to the substructures discussed in \cite{Chou2007} and \citet{Pila2014}, it is curious whether they are relevant with each other. 

In this work, we use the M giant stars observed from the LAMOST survey to revisit the less prominent substructures in the region of the north Sgr leading tail. As of June 2014, the LAMOST survey has already obtained more than 4 million stellar spectra with sufficient signal-to-noise ratio (S/N). \citet{ZJ15} developed a template-matching technique to reliably identify the M giant stars from the low resolution spectra.
Applying their method to the LAMOST DR2 data, we obtain 17,589 M giant stars, which is so far the largest sample of the M giant spectra. On the other hand, \citet[hearafter ~\li]{2016arXiv160300262L}, employed a photometric-based method to estimate [Fe/H] and distance for the M giant stars.

The paper is organized as following. In section 2, we describe how to select the data and improve the estimation of the metallicity. In section 3, the substructures in the region of the Sgr leading tail is unveiled and featured in spatial-kinematic-metallicity space. In section 4, we discuss the possible origins of the new unveiled substructures. Finally, a brief conclusion is drawn in the last section.

\section{Data}

\subsection{The updated LAMOST M giant star catalog}
The Large Sky Area Multi-Object Fiber Spectroscopic Telescope (LAMOST; also called Guo Shou Jing Telescope) is a National Key Scientific facility built by the Chinese Academy of Sciences \citep{cui2012,zhao2012}. The LAMOST spectroscopic survey, started since 2011, mainly aimed for the understanding of the structure of the Milky Way \citep{2012RAA....12..735D}. Although the standard data processing pipeline provides quite accurate estimation of the stellar parameters for the FGK type stars, it does not reliably identify the M type stars \citep{Luo2015}. In order to identify the M type stars and classify the M giant and dwarf stars from the LAMOST dataset,
 \citet{ZJ15} constructed the M-giant templates and identified around 9,000 M-giant stars from LAMOST DR1 dataset. We apply the same method to the LAMOST DR2 dataset and expand the M giant sample to 21,696. Then we purify the sample by excluding a few contaminated K giant stars and M dwarf stars using the criteria of the WISE color index \ww\ according to~\li\ and \citet{ZJ15}, finally we get 17,589 M giant stars. The interstellar reddening is corrected using the same spatial model of the extinction mentioned in~\li. We adopt the $E(B-V)$ maps of \citet{Schlegel1998}, in combination with  $A_r/E(B-V) = 2.285$ from \citet{Schlafly2011} and $A_\lambda/A(r)$ from \citet{Davenport2014} but has a latitude dependence. 
 
 The contamination of 166 carbon stars have also been excluded by cross-matching with the latest LAMOST carbon star catalog (Ji et al. in preparation).

\subsection{Distance and metallicity determination}
\begin{figure*}
\centering
%\figurenum{1}
\includegraphics[scale=0.60,angle=0]{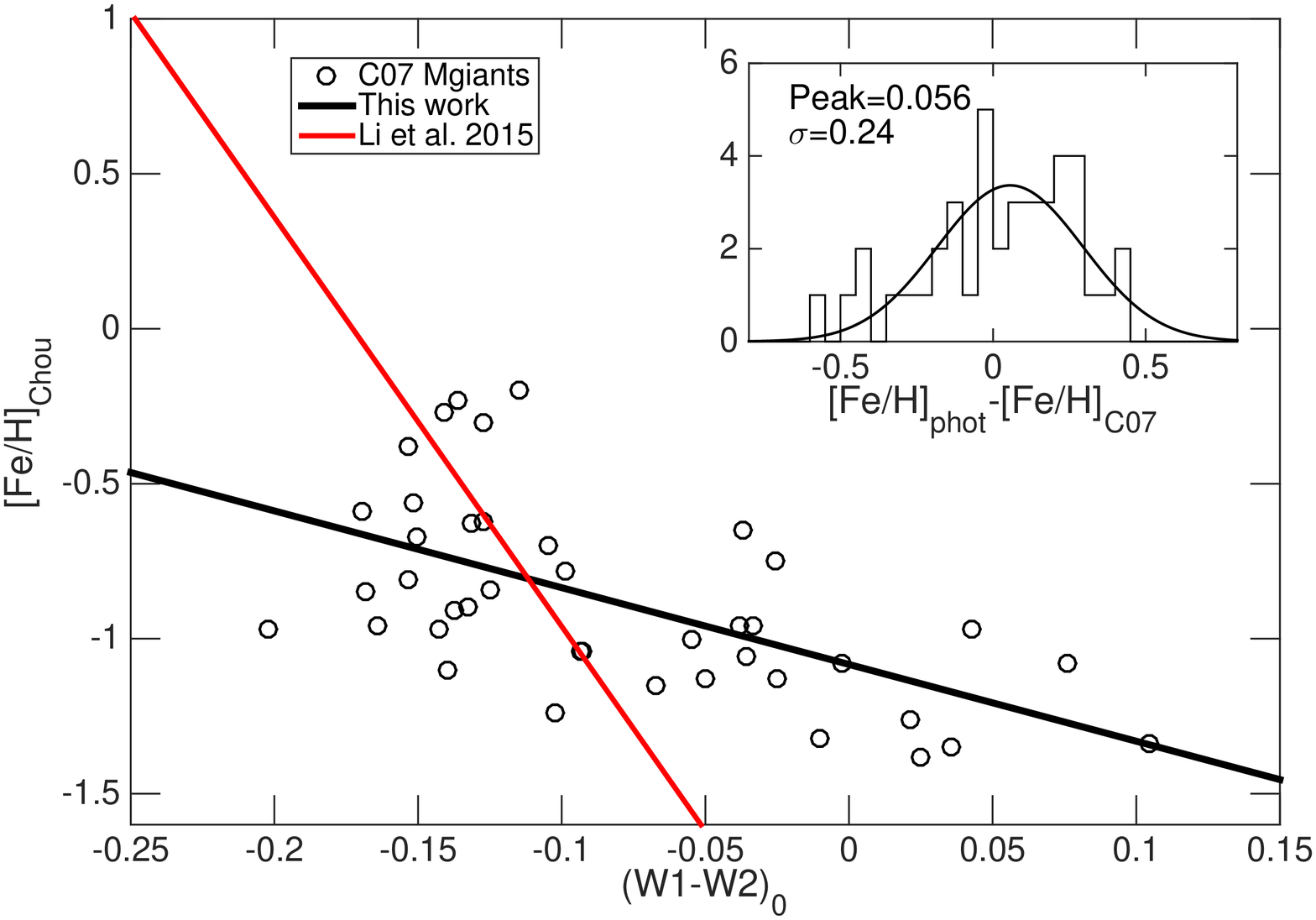}
\caption{The metallicity vs. \ww\ of the 41 M giant stars from \chou. The black line shows the best-fit linear relationship in this work. The red line shows the linear relationship for APOGEE M giant stars from \li. The inset histogram shows the scatter of metallicity residual, which has a dispersion of 0.24 dex. The mean standard deviation of the abundance determined by \chou\ is 0.086\,dex and the mean error of \ww\ is 0.031 mag.}
\label{fig:feh}
\end{figure*}

\li\ developed a photometric-based method to estimate the distance for the M giant stars. Specifically, the absolute magnitude in $J$ band is determined from
\begin{equation}
\rm M_{J}=3.12[(J-K)_{0}^{-2.6}-1]-4.61.
\label{eq:distance}
\end{equation}
For the LAMOST M giant sample used in this work, we adopt the equation (\ref{eq:distance}) to estimate the distance. Based on the high-resolution spectroscopic data, \chou\ presented metallicities for 59 M giant stars, which are the member candidates for the Sgr stream according to their positions and velocities.  We use this sample to improve the photometric metallicity derived by~\li. We apply the photometric cut mentioned in section 3.1 of~\li\ to the \chou\ sample and obtain 41 stars with high quality in photometry. The mean error of $W_1-W_2$ for these stars is only 0.031 mag. Figure~\ref{fig:feh} shows the metallicities of the 41 M giant stars derived by \chou\ as a function of \ww. Then, the correlation between [Fe/H] and \ww\ is fitted with the following linear relation:
\begin{equation}
{\rm [Fe/H]}_{\rm WISE}=-2.477\times(W_1-W_2)_{0} -1.083,
\label{eq:metallicity}
\end{equation}
and shown with the black line in the figure. The residual scatter of the linear relation, as shown in the inset of Figure~\ref{fig:feh}, is 0.24\,dex. As a comparison, the red line shows the linear relationship obtained from the APOGEE M giant stars in~\li. It is obvious that the linear relationship between \ww\ and [Fe/H] in~\li\ does not well fit the data from \chou. 
In~\li, the selection of the M giant stars is purely dependent on the photometry (see their section 3.3), which may be polluted by the K giant contaminations. This difference also could be caused by population effect. Considering the large fraction of K giant stars in the APOGEE data, this contamination may not be ignored and hence, to some extent, induce systematic bias in the [Fe/H]--\ww\ correlation. For the M giant sample from \chou, the M giant identification has been confirmed from the high-resolution spectra and thus should not be affected by the K giant stars. Therefore, the updated linear relationship based on the \chou\ sample (the black line) should be more reliable. In the rest of this paper, we use this improved metallicity estimates for the LAMOST M giant stars. 

\section{Results}\label{sect:result}
In order to better compare with the Sgr stream, we adopt the Sgr coordinates originally introduced by \citet{Majewski2003}. The equator of the coordinate system is defined by the mid-plane of the Sgr stream. The latitude \Bsgr\ of the Sgr coordinates is parallel with the mid-plane of the stream as viewed from the Sun. The zero-point of the longitude \Lsgr\ is at the core of the Sgr dwarf galaxy and \Lsgr\ increases toward the direction of the trailing tidal tail. The coordinates with \Lsgr$<180^\circ$ are mostly located in the south of the Galactic disk mid-plane and the coordinates with \Lsgr$>180^\circ$ are mostly in the north.

The line-of-sight velocity measured from the M giant spectra is with respect to the Sun. We convert it to \vgsr, the line-of-sight velocity with respect to the Galactic standard of rest, using the following equation:
\begin{equation}
V_{gsr}=V_{los}+10{\rm cos}l{\rm cos}b+225.2{\rm sin}l{\rm cos}b+7.2{\rm sin}b,
\end{equation}
in according to~\citep{2012ApJ...757..151L,1998MNRAS.298..387D}.

Figure~\ref{fig:Mgiant_full_lambda} shows the line-of-sight velocities (top panel) and distances (bottom panel) along \Lsgr\ for the LAMOST DR2 M giant stars. The green circles show all the M giant stars located with $-15^\circ<$\Bsgr$<15^\circ$ and heliocentric distance larger than $10$\,kpc. The red lines indicate the Sgr stream from \citet[][hereafter B14]{Belokurov2014}, who marked them with sub-giant, red giant branch, and blue horizontal branch stars. And the grey dots are the simulation data containing the last wrap of the stream from \lm. The black diamonds are from the substructures of \chou, with the distance estimated using the same way as in~\li. In the region of $90^\circ<$\Lsgr$<150^\circ$, which targets the trailing tail, the M giant stars fit quite well in both velocity and distance with either the simulation data from \lm\ or the observed data from \belokurov. The clumpy data located at $130^\circ<$\Lsgr$<200^\circ$ and \vgsr$\sim0$\,\kms are mostly the contamination from the disk population since this region crosses through the Galactic disk in the anti-center direction. However, in the top panel, it is noted that a narrow tail located within $130^\circ<$\Lsgr$<150^\circ$ with quite small velocity dispersion is relatively isolated out of the clumpy disk contaminations. These stars are likely members of the leading tail.

In the region of $200^\circ<$\Lsgr$<320^\circ$, the stars are separated into two groups in distance. The one with distance larger than 20\,kpc is consistent with both \lm\ and \belokurov\ in distance and \vgsr\ and hence belongs to the Sgr leading tail. The other stars located between 10 and 20\,kpc in distance is not identified by \belokurov\ and neither the distance nor the velocity is in agreement with the simulation data from \lm. Some of the stars in this group are well overlapped with the Sgr leading tail in velocity as shown in the top panel of figure~\ref{fig:Mgiant_full_lambda}. Some stars are well overlapped with the Sgr leading tail in velocity are from substructure of \chou. A few stars shift from the Sgr leading tail by about 100\kms toward the positive velocity are NGC group of \chou. Therefore, it seems that the stars with distance at 10--20\,kpc may be further separated into two groups, one with similar \Lsgr--\vgsr\ trend as the Sgr leading tail, but located in much nearer distance, the other with larger \vgsr\ than the Sgr leading tail by $\sim100$\,\kms\ at the same \Lsgr\ and overlapped with the \ngc\ group from \chou.

The LAMOST M giant stars well indicate the Sgr tidal stream in both north and south tails. This can therefore better constrain the dynamics of the tidal stream. However, before addressing the orbital properties of the Sgr tidal stream, it is very important to clarify whether those not prominent but clearly displayed substructures, i.e. the two groups of stars located within 10--20\,kpc, are related with the tidal stream. 

\subsection{Substructures in $210^\circ<$\Lsgr$<290^\circ$}
\begin{figure*}
\centering
%\figurenum{2}
\includegraphics[scale=0.50,angle=0]{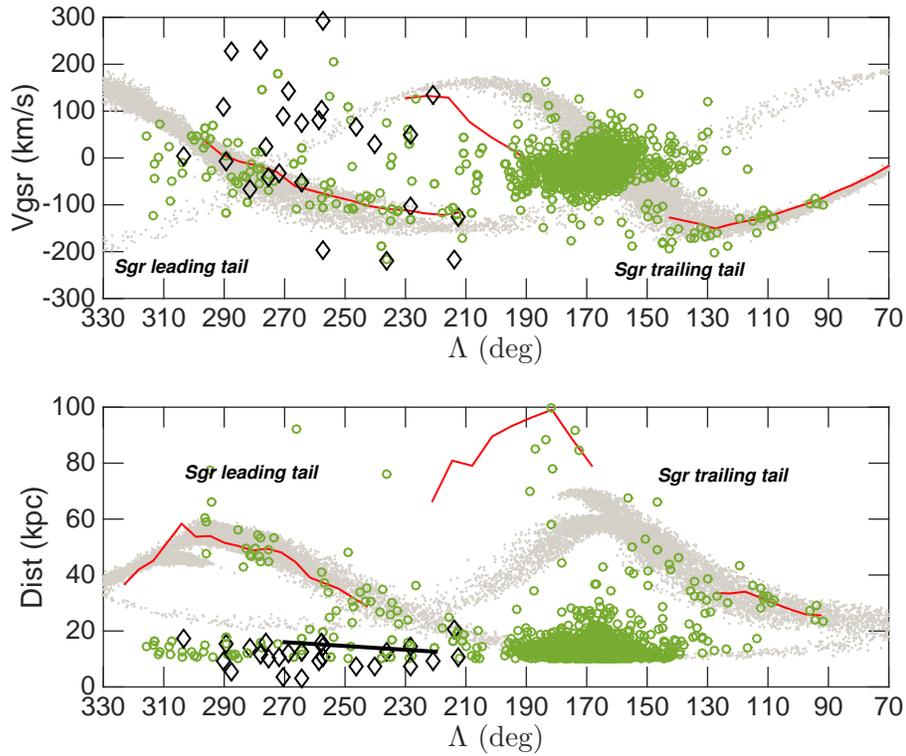}
\caption{Line-of-sight velocity vs. \Lsgr\ (top panel) and distance vs. \Lsgr\ (bottom panel) for the M giant stars with $15^\circ<$\Bsgr$<15^\circ$. The green circles show the M giant stars with distance larger than 10 kpc. The black diamonds show NGC group and the near Sgr sample from \chou. The red line show the detections from \belokurov, who derived them from the sub-giant, red giant branch and blue horizontal branch stars. The grey points show the simulation data from \lm. The black line shows the location of the unknown over-density from \citet{heidi07}.}
\label{fig:Mgiant_full_lambda}
\end{figure*}

In order to further investigate the confusing velocity substructures shown in figure~\ref{fig:Mgiant_full_lambda}, we select the stars within $210^\circ<$\Lsgr$<290^\circ$ to avoid possible contaminations from the thick disk and the some over-density. Indeed, the stars located at $290^\circ<$\Lsgr$<320^\circ$ well overlap with the RR Lyrae substructure discovered by~\cite{Duffau2014}. At the other end, the region of $200^\circ<$\Lsgr$<210^\circ$ is very close to the Galactic disk in the Galactic anti-center region. The rest of the region well overlaps the Sgr leading tail in the longitude. The stars in interest are located between 10 and 20\,kpc, while the Sgr leading tail is well identified at the distance between 20 and 60 kpc. 

As seen from the top panel of figure~\ref{fig:Mgiant_full_lambda}, both the stars in interest and the distance-identified Sgr leading tail contribute to the group with velocity from $\sim0$\,\kms\ at \Lsgr$\sim290^\circ$ to $\sim-150$\,\kms\ at \Lsgr$\sim210^\circ$. However, only the stars located between 10 and 20 kpc contribute to the group with the velocity from $\sim+200$\,\kms\ at \Lsgr$\sim290^\circ$ to $0$\,\kms\ at \Lsgr$\sim210^\circ$. A apparent concern is that the group of stars with larger velocity are possibly contaminated by the thick disk. Therefore, we firstly investigate the distribution of the metallicity for these stars and remove those with high possibility to be the thick disk stars. 

\begin{figure}
\centering
%\figurenum{4}
\includegraphics[scale=0.60,angle=0]{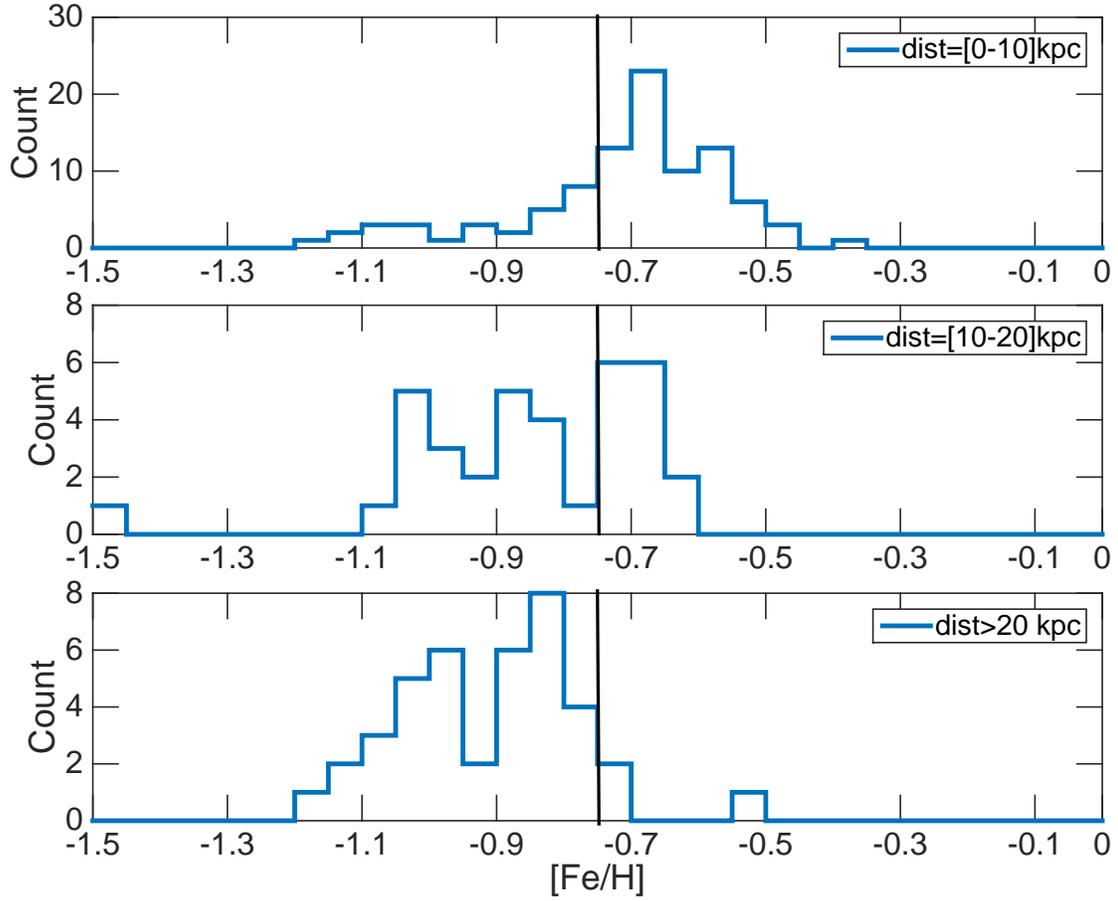}
\caption{[Fe/H] distribution in different range of distance. The blue lines indicate the metallicity distribution for the M giant stars with distance smaller than 10\,kpc, between 10 and 20\,kpc, and beyond 20\,kpc in the top, middle, and bottom panels, respectively. The vertical lines, which are located at [Fe/H]$=-0.75$\,dex, indicate the point at which we separate the halo stars from the thick disk ones. Those with [Fe/H] smaller than -0.75\,dex are not severely contaminated by the thick disk and selected for the detection of the substructures.}
\label{fig:feh_compare}
\end{figure}

Figure~\ref{fig:feh_compare} shows the metallicity distribution for the M giant stars in three ranges of distances, 0--10\,kpc (top panel), 10--20\,kpc (middle panel), and beyond $20$\,kpc (bottom panel). In principle, the thick disk stars should dominate the metallicity distribution function for the nearest stars. Indeed, the peak of the metallicity distribution for the stars within 10\,kpc in distance is at around -0.7\,dex, well consistent with the thick disk populations. In the distance larger than 20\,kpc (the bottom panel), a metal-poor population dominates the metallicity distribution with almost no stars fell in [Fe/H]$>-0.75$\,dex. In the distance between 10 and 20\,kpc, the metallicity distribution function seems a mixture of the thick disk stars with typical metallicity distribution as shown in the top panel and the metal-poor population represented by the bottom panel. If we select the M giant stars with [Fe/H]$<-0.75$\,dex in the distance between 10 and 20\,kpc, we can significantly reduce the contaminations from the thick disk. Assume that the stars within 10 kpc are all from the thick disk. This is obviously a quite crucial assumption. Then we can read from the top panel that about 20\% stars are within [Fe/H]$<-0.75$\,dex. Given that in the middle panel, all stars with [Fe/H]$>-0.75$\,dex are from the thick disk, the possible contamination from the thick disk in the region of [Fe/H]$<-0.75$\,dex is only about 15\%, according to the star counts in the panel. Therefore, cutting out the stars with [Fe/H]$>-0.75$\,dex can effectively avoid the effect of the thick disk in velocity.

\begin{figure}
\centering
%\figurenum{5}
\includegraphics[scale=0.50,angle=0]{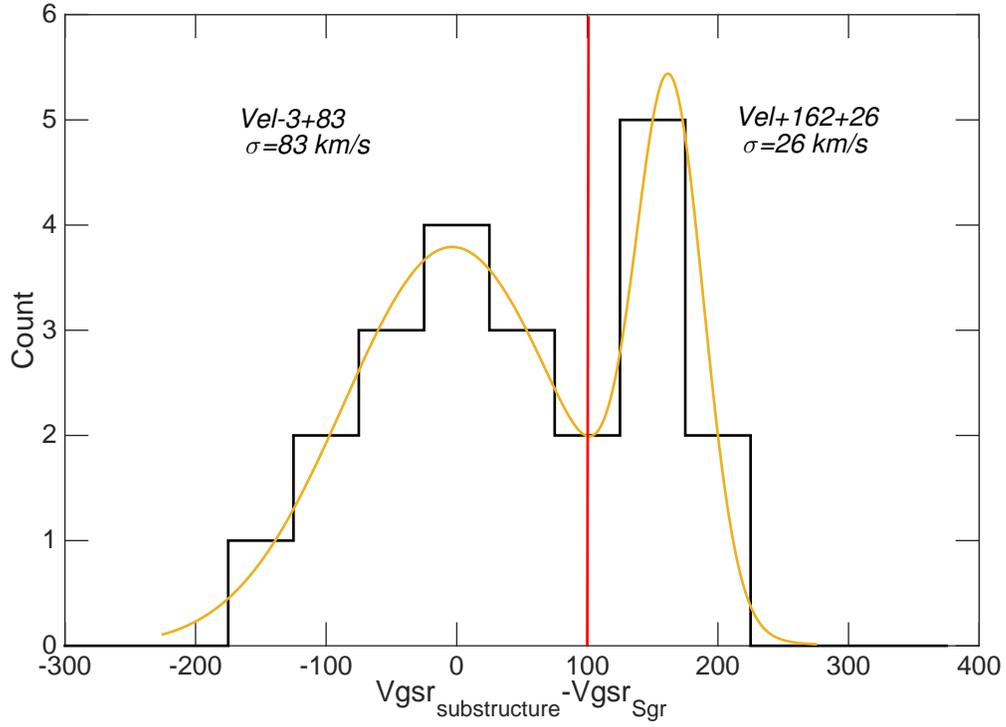}
\caption{ The black line indicates the distribution of the \vgsr\ offset with respect to the Sgr leading tail at the same \Lsgr\ for the selected M giant stars with [Fe/H]$<-0.75$, $-15^\circ<$\Bsgr$<15^\circ$, and $210^\circ<$\Lsgr$<290^\circ$. The reference velocity of the Sgr leading tail is from \belokurov. The yellow line is the best fit with two Gaussians. The red vertical line is the separation between \structA\ (left) and \structB\ (right).}
\label{fig:structure_vgsr_hist}
\end{figure}

\begin{figure}
\centering
%\figurenum{6}
\includegraphics[scale=0.40,angle=0]{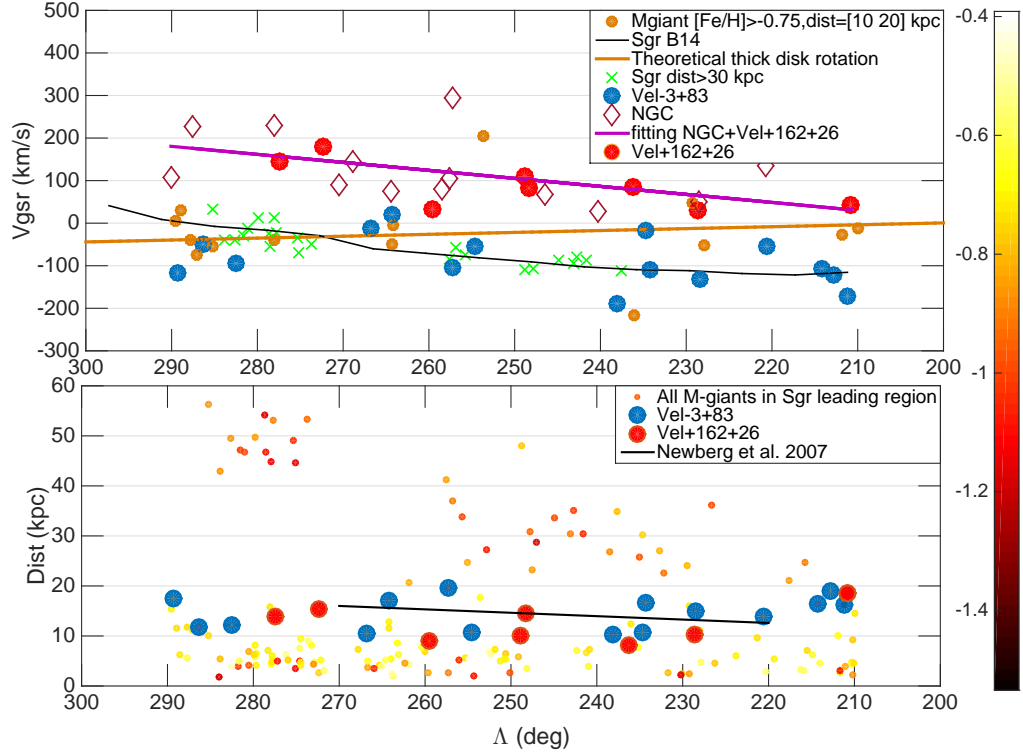}
\caption{The distribution in the \vgsr\ vs. \Lsgr\ plane (the top panel) and the distribution in the distance vs. \Lsgr\ plane (the bottom panel) for the two substructures, \structA\ and \structB. The blue filled circles show the member stars of \structA\ and the green crosses are the member stars of the Sgr leading tail located beyond 30\,kpc in distance. The black line in the top panel indicates the velocity of the Sgr leading tail derived by \belokurov. The red filled circles are the member stars of \structB\ and the magenta diamonds are the stars from the \ngc\ group in \chou. The magenta line indicates the linear fit for the combined data of \structB\ and the \ngc\ group stars from \chou. The yellow dots are the stars with distance between 10 and 20\,kpc and [Fe/H]$>-0.75$,dex. The yellow line indicates the velocity of the thick disk according to eq.~(\ref{eq:disk_velocity}). The black line in the bottom panel displays the location of the unknown over-density at (\Lsgr,$g_0$)=($240^\circ$, 16.7) from \citet{heidi07}. The color bar indicate the metallicity for the lower panel small dots.}
\label{fig:vgsr_compare1}
\end{figure}

We then quantify the distribution of \vgsr\ for the M giant stars with $210^\circ<$\Lsgr$<290^\circ$ and [Fe/H]$<-0.75$\,dex. Because the velocity \vgsr\ is correlated with \Lsgr, it is not easy to directly derive the velocity distribution. To solve this issue, we select the velocity of the Sgr leading tail as the baseline, and derive the distribution of the velocity offset from the baseline at a given \Lsgr\ for the selected stars. We adopt the velocity trend of the Sgr leading tail from \belokurov\ (the red line in Figure~\ref{fig:Mgiant_full_lambda}) as the baseline. Figure~\ref{fig:structure_vgsr_hist} shows the distribution of the velocity offset (black line) with the bin size of $50$\,\kms. The distribution shows clear bimodality, one peak is located at around 0\,\kms, while another narrower peak is located at about 150\,\kms. In order to investigate whether the bimodality is real or just statistical fluctuation, we calculate the Akaiki?s information criterion (AIC) and Bayesian information criterion (BIC) for the mixed Gaussian models with various components. We find the BIC (AIC) for 1-, 2-, and 3-Gaussian models are 16.5 (15.6), 11.7 (10.2), and 18.6 (16.2), respectively. According to the definition of BIC and AIC, the model with minimum value is the best model for the data. This suggests that a two-Gaussian model is the best one to fit the data.
 A two-Gaussian model is then fitted to the distribution (yellow line) and indicates that the best fit peaks are $-3$\,\kms\ with velocity dispersion $83$\,\kms and $+162$\,\kms\ with velocity dispersion of $26$\,\kms, respectively. According to the bimodal fitting, we separate the stars into two groups at $V_{gsr}-V{gsr,Sgr}=100$\,\kms\ and display their \Lsgr--distance--\vgsr\ distribution in Figure~\ref{fig:vgsr_compare1}. The stars with velocity offset smaller than $100$\,\kms\ are denoted as \structA\ and marked as blue filled circles and those with offset larger than $100$\,\kms\ are denoted as the \structB\ and marked as red filled circles. For better comparison, Figure~\ref{fig:vgsr_compare1} also indicates the NGC group from \chou\ (red diamonds), the possible thick disk M giant stars with [Fe/H]$>-0.75$\,dex in 10--20\,kpc(yellow dots), and the M giant member candidates of the Sgr leading tail located beyond 30\,kpc (green crosses). 

First, the velocity dispersion for the substructure~\structA\ is $83$\,\kms, which is consistent with the large velocity dispersion shown with the red points in figure 2 of \chou. The correlation between \vgsr\ and \Lsgr\ is also quite similar to the sample from \chou. Second, although the trend of \vgsr\ along with \Lsgr\ is also comparable with the Sgr leading tail beyond 30\,kpc in distance, the velocity dispersion measured from our sample is significantly larger than the Sgr leading tail, whose dispersion is only $36$\,\kms\ (but see further discussion at section~\ref{sect:disc_structA}). The M giant stars belonging to this group are listed in table~\ref{tab:nearbybranch}.

Second, the \ngc\ group stars from \chou\ are located at exactly the same velocity as the substructure~\structB. Considering that the two groups of stars are also at similar distance as shown in the bottom panel of Figure~\ref{fig:Mgiant_full_lambda}, they are likely from the same substructure. The purple straight line shows the linear fitting of \vgsr\ along \Lsgr\ for the substructure \structB\ using the data both from this work and \chou. Because the \ngc\ group stars in \chou\ are actually start from 5\,kpc, two more stars located slightly smaller than 10\,kpc in the same sky area and velocity regime are added to \structB. The candidates of the~\structB\ are listed in table~\ref{tab:NGC}. 

Finally, we present that the metal-rich M giant stars (yellow dots) do have similar kinematics to the thick disk. For this purpose, we build a oversimplified thick disk kinematical model with azimuthal velocity of $170$\,\kms\ and naively assume that all the stars are located at 15 kpc from the Sun. We calculate the line-of-sight velocity at various \Lsgr\ for the thick disk in the following way \citep{2012ApJ...757..151L}:
\begin{eqnarray}
 V_{td}=-{{\sin l}\over{|\sin l |}}V_{rot}\sin a\cos b,\nonumber\\
r= \sqrt{d^2+8^2-2 \times8 \times d \times \cos l},\\
 a=\arccos({{d^2+r^2-8^2}\over{2dr}}).\nonumber
\label{eq:disk_velocity}
\end{eqnarray}
where $8$\,kpc is the adopted distance from the Sun to the Galactic center, $d$ is the distance to the observed star (in this case $15$\,kpc), $r$ is the distance from the Galactic center, $a$ is the azimuth angle of the observed star with respect to the Sun--Galactic center line, $V_{rot}$ is the adopted azimuthal velocity of the thick disk ($170$\,\kms), and $V_{td}$ is the predicted line-of-sight velocity with respect to the Galactic standard of rest. The yellow line in Figure~\ref{fig:vgsr_compare1} shows the predicted line-of-sight velocity for the thick disk. The metal-rich stars within 10--20\,kpc (yellow dots) are well consistent with it, implying that most of the stars are exactly the thick disk stars, except 2 of them located beyond $200$\,\kms. Moreover, neither the substructure~\structA\ nor \structB\ show any similarity to the model velocity of the thick disk. This again confirms that both the substructures are not likely contributed by the thick disk.

\section{Discussion}\label{sect:discussion}
In this section we clarify whether the two substructures located in $210^\circ<$\Lsgr$<290^\circ$ and distance between 10 and 20\,kpc are related to the Sgr tidal stream.

\subsection{\structA}\label{sect:disc_structA}
\begin{figure}
\centering
%\figurenum{7}
\includegraphics[scale=0.50,angle=0]{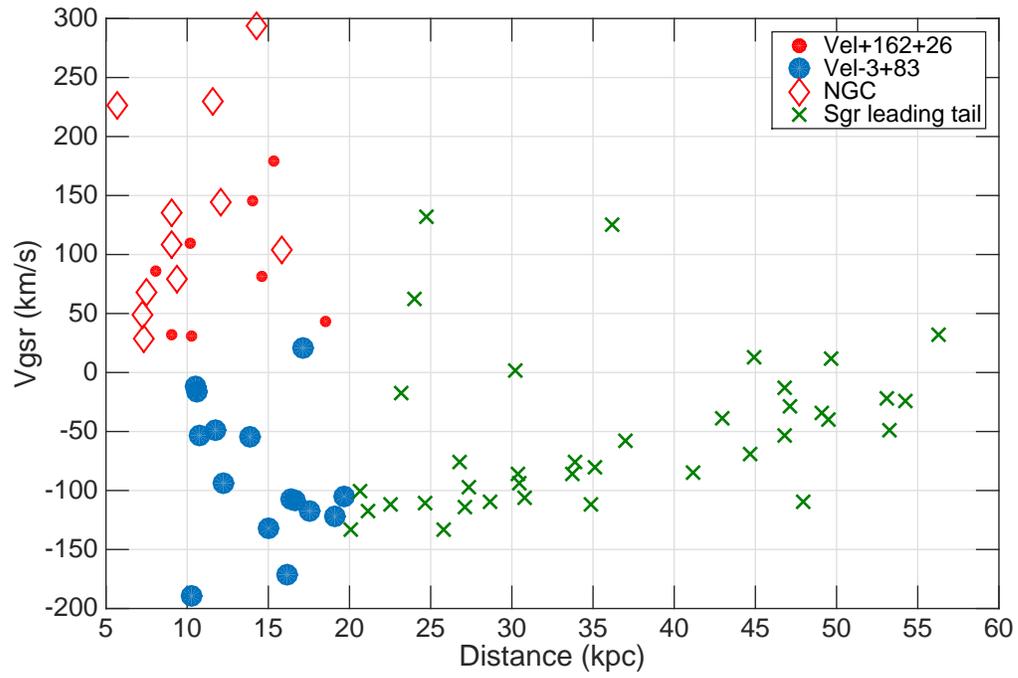}
\caption{Vgsr vs. distance distribution for the Sgr leading tail, \structA, and \structB. The symbols are same as in figure~\ref{fig:vgsr_compare1}.}
\label{fig:dist_vgsr1}
\end{figure}

The substructure \structA\ shows very similar velocity to the Sgr leading tail but the velocity dispersion is quite larger than the Sgr leading tail. However, when the member stars of \structA\ are mapped to the \vgsr\ vs. distance plane (see figure~\ref{fig:dist_vgsr1}), we find that the large velocity dispersion of the \structA, shown as the blue filled circles, is due to the rapid variation of \vgsr\ when distance changes. The actual velocity dispersion can be derived from the velocity scattering at each given distance bin. We then obtain that the velocity dispersion is only 30.4\,\kms, quite comparable with the stars in the Sgr leading tail (green crosses). Moreover, the substructure \structA\ seems continuously connected with the Sgr leading tail at the distance of 20\,kpc. However, putting \structA\ and the leading tail in 3-dimensional space, the connection at 20\,kpc produces a folding 3D structure, which should not be a realistic tidal stream. 

\begin{figure*}
\centering
\begin{minipage}{18cm}
\centering
\includegraphics[scale=0.5]{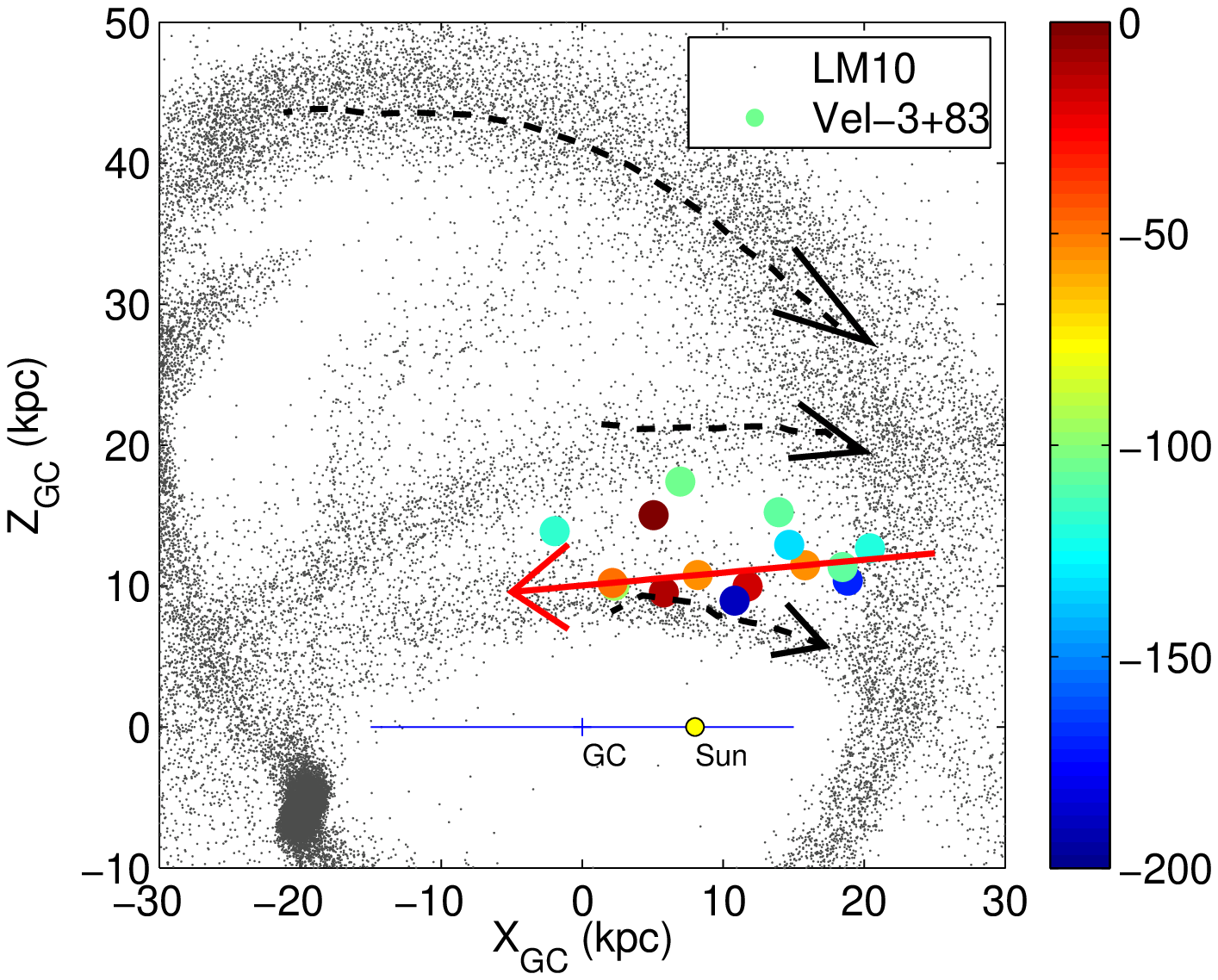}
\includegraphics[scale=0.5]{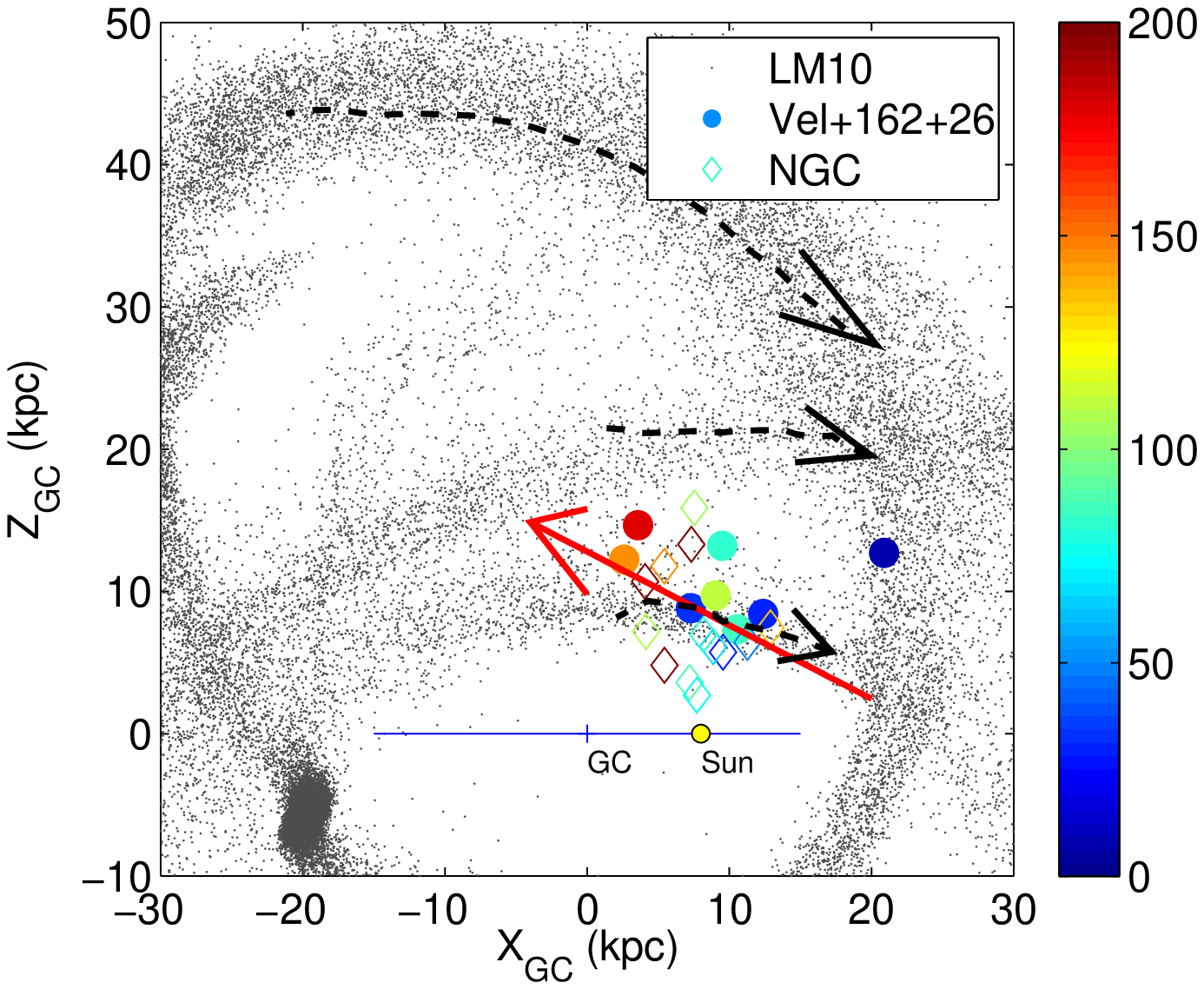}
\end{minipage}
\caption{Left panel: The XZ map of the substructure \structA. The color dots are the member stars of \structA\ with the color coded \vgsr. The red arrow line indicates the approximate direction of motion of the substructure. The gray dots are the simulation data from \lm. The dashed arrow lines indicate the directions of motion for three tidal tails of the simulation located nearby \structA. The blue line stands for the Galactic disk mid-plane, the short vertical stroke in the center of the blue line represents the Galactic center, and the yellow dots stands for the location of the Sun. 
Right panel: The XZ map of the substructure \structB\ and the \ngc\ group of \chou. The color dots are the member stars of \structB\ with color coded \vgsr, while the hollow color diamonds represent the \ngc\ group stars. Notice that the color scale in the right panel is different with that in the left panel. The red line indicates the rough direction of motion of the stars from both \structB\ and the NGC group. The gray dots and the dashed arrow lines are same as those in the left panel.}\label{fig:branch_cartoon}
\end{figure*}

According to the \Lsgr--distance--\vgsr\ relation, we are able to constrain the direction of motion of the substructure~\structA. \structA\ member stars (color dots) and the Sgr leading tails (gray dots) from \lm\ are plotted in $X$--$Z$ plane in the left panel of figure~\ref{fig:branch_cartoon}. The colors of the \structA\ stars indicate \vgsr. It shows that at around ($X$, $Z$)=(20, 10)\,kpc, \vgsr\ of the \structA\  is toward the Sun, implying that the substructure \structA\ is moving to negative  $X$ direction. At around ($X$, $Z$)=(5, 10)\,kpc, \vgsr\ is roughly zero, meaning that at this point the substructure is roughly moving along the tangental direction with respect to the line of sight. Combine these information together, we can infer that \structA\ is likely moving from right to left in the $X$--$Z$ plane (toward the inner Galaxy), showing as a red arrow in the figure. 
%Assuming that the substructure is a stream and its spatial velocity at each point is tangential to the stream. Then if at one point the line-of-sight velocity, \vgsr, is negative, it means that the three-dimensional velocity is toward the Sun at the point and vice versa. From the \Lsgr--distance--\vgsr\ of the member stars, we infer that, at \Lsgr$\sim210^\circ$, the stars is moving toward the Sun and at \Lsgr$\sim290^\circ$, \vgsr$\sim0$, which means that the stars are roughly moving in the perpendicular direction with respect to the line-of-sight. Combining these together, we can roughly point out that the direction of motion of the substructure~\structA\ is toward the inner galaxy, as shown with the red arrow line in the left panel of figure~\ref{fig:branch_cartoon}. Figure~\ref{fig:branch_cartoon} also demonstrates the direction of motion in the X vs. Z plane for the Sgr tidal tails based on the simulated data from \lm. 
It is obvious that the direction of motion for \structA\ is opposite to any of the Sgr tidal tails. Hence, it is very difficult to attribute the substructure to a part or branch of the Sgr tidal tail. Alternatively, the substructure \structA\ could be a disrupting satellite of the Sgr dwarf galaxy. However, the distance from the \structA\ to the leading tail is as large as 40\,kpc, which seems too far to be a satellite of the Sgr dwarf galaxy. Therefore, we infer that it is very likely a new tidal debris not related to the Sgr dwarf galaxy.

It is noted that~\citet{heidi07} discovered an unknown over-density from the BHB stars at (\Lsgr, $g_0$)=(240$^\circ$, 16.7). \structA\ is exactly located at the same location of the over-density (see the comparison in the bottom panel of figure~\ref{fig:vgsr_compare1}). Table~\ref{tab:nearbybranch} also shows that most of the member M giant stars of \structA\ are located above the mid-plane of the Sgr tidal stream (\Bsgr$>0^\circ$), in agreement with the figure 3 in \citet{heidi07}.

It is also worthy to note that the ambiguous arm found by \citet{Pila2014}, who claimed that it is a new wrap of the Sgr stream without velocity information, covers the same sky area as the substructure~\structA. Therefore, it seems that what the authors found is not part of the Sgr tidal debris, but the same tidal debris unveiled in this work.

\subsection{\structB}
Figure~\ref{fig:vgsr_compare1} also displays the member stars of the substructure~\structB\ with red filled circles combined with the NGC group members from \chou\ with red diamonds. \chou\ argued that these stars are likely the dynamically old Sgr members from the wrapped trailing tail. Similar to \structA\, we can also estimate the approximate direction of motion for \structB\ using the relationship of \Lsgr--distance--\vgsr\ and find that \structB\ is coarsely moving toward smaller $X$ and larger $Z$, as shown in the right panel of figure~\ref{fig:branch_cartoon}. This direction of motion is opposite to all the Sgr tidal streams predicted by the simulation data from \lm\ and thus seems against the statement by \chou. Therefore, we propose that this substructure is also likely a new tidal debris not relating to the Sgr tidal stream.

\section{CONCLUSION}

Using 17,000 M giant stars from the LAMOST DR2, we are able to map the Sgr tidal stream in the whole northern sky. Both the leading tail and the trailing tail are well sampled. These data will be very important to constrain the orbit of the Sgr tidal stream. Before addressing the orbital properties of the Sgr tidal stream, we investigate whether there is any more weaker and less prominent substructures nearby the Sgr tidal stream and if there is any, whether it belongs to the Sgr. 

We find two substructures in the north Galactic cap region around the Sgr leading tail. The substructure~\structA\ shows very similar \Lsgr--\vgsr\ relation to the leading tail with nearer distance between 10 and 20\,kpc. However, based on the analysis of the spatial position and the possible direction of motion, it is very likely a new tidal debris not belonging to the Sgr leading tail, nor the trailing tail in north. We point out that the spatial position of \structA\ is consistent with the unknown over-density discovered by \citet{heidi07}.

The other substructure~\structB\ shows positive \vgsr\ in the similar range of distance. It well overlaps with the \ngc\ group found by \chou, who attribute it to the earlier wrapped Sgr trailing tail. However, when we map the stars from both \structB\ and the NGC group of \chou\ to the $X$--$Z$ plane, we find that the direction of motion of this combined substructure is also not in consistent with any branch of the Sgr tidal stream. Therefore, we propose that this is another weak tidal debris not relating to the Sgr tidal stream.

Our proposal for the substructure \structA\ and \structB\ is based on that the simulation of the Sgr tidal process from \lm\ is relatively correct in the distance and velocity. However, if the simulation has bias from the real Sgr tidal stream, which is still far from completely observed, then our conclusion has to be revisited.

In this work, we also revise the relationship between metallicity and WISE color index \ww, using 41 M giant stars with reliable metallicity estimates from the high resolution spectra (\chou).

In the next two years, LAMOST will finish its survey and we expect that it will expand the M giant sample by a factor of 2. Then, we will have much more sample to better address the two substructure and give tighter constraints on their origins.

\section{ACKNOWLEDGEMNTS}

We thank Martin Smith for his kind support on this project.
 This work is supported by the Strategic Priority Research Program ``The Emergence of Cosmological Structures" of the Chinese Academy of Sciences grant No. XDB09000000 and the National Key Basic Research Program of China 2014CB845700. C.L. acknowledges the National Natural Science Foundation of China (NSFC) under grants 11373032, 11333003, and U1231119. 
J.L, J.L.H and J.Z. thank the NSFC under grants 11173044, 11503066 and the Shanghai Natural Science Foundation 14ZR1446900.Guoshoujing Telescope (the Large Sky Area Multi- Object Fiber Spectroscopic Telescope LAMOST) is a National Major Scientific Project built by the Chinese Academy of Sciences. Funding for the project has been provided by the National Development and Reform Commission. LAMOST is operated and managed by the National Astronomical Observatories, Chinese Academy of Sciences.

\clearpage

\clearpage

\clearpage

\clearpage

\clearpage

\clearpage

\clearpage

\clearpage
\begin{table}
\begin{center}
\caption{Positions, [Fe/H], \vgsr, and distance for the candidates of \structA. The [Fe/H] and distance are photometrically determined.}
\label{tab:nearbybranch}
\begin{tabular}{lcccccccccc}
\hline
~ & RA & Dec & $\Lambda$ &Beta& [Fe/H]$_{phot}$ & \vgsr\ & dist$_{phot}$ \ & J$_{0}$\\
~& deg & deg & deg &deg& dex & \kms\ & kpc \ & mag\\
\hline
1&138.3828 & 19.5730 & 211.1521 &9.9894& -1.53 & -171.70& 16.16 & 11.55\\
2&140.1950 & 19.8632 & 212.8254 &9.4654& -0.87 & -121.57 & 19.07 & 10.14\\
3&141.9523 & 21.4288 & 214.2254 &7.6658& -1.07 & -107.03 & 16.42 & 10.03 \\
4&154.3984 & 40.5305 & 220.5967 &13.2235& -0.91 & -54.58 & 13.86 & 10.47\\
5&158.6647 & 24.8682 & 228.4452 & 0.8282&-0.99 & -131.70 & 15.00 & 10.66\\
6&165.8651 & 25.9625 & 234.2687 & 2.2673&-0.90 & -108.51 & 16.67 & 9.91\\
7&171.5664 & 37.1245 & 234.7326 & -14.4256&-0.86 & -16.74 & 10.61 & 11.26\\
8&165.2152 &13.0378 &238.1064 &10.0947& -1.00& -189.21 & 10.26 & 10.26\\
9&189.5757 & 24.2840 & 254.6262 &-9.5470& -0.91 & -53.63 & 10.78 & 10.58\\
10&180.8900 & 1.7851 & 257.2424 &14.3275& -1.02 &-105.37 & 19.65 & 11.47\\
11&186.8714 & -1.4641 & 264.2678 &14.5039& -0.97 & 20.97 & 17.17 & 11.40\\
12&191.7561 & 2.4719 & 266.7687 &8.7293& -0.89 & -12.08 & 10.48 & 10.59\\
13&204.5508 & -6.4734 &	282.5108 &10.1858& -0.87 & -93.78 & 12.27 & 9.00\\
14&218.5212 & 9.6072 & 286.3416	&-10.7381& -1.06 & -49.02 & 11.75 & 11.35\\
15&213.1342 & -5.3586 &	289.3505 &4.8765& -0.83 & -117.16 & 17.59 & 10.96\\

\hline
\end{tabular}
\end{center}
\end{table}

\begin{table}
\begin{center}
\caption{Positions, [Fe/H], \vgsr, and distance for the candidates of \structB.}
\label{tab:NGC}
\begin{tabular}{lcccccccccc}
\hline
~ & RA & Dec & $\Lambda$ &Beta& [Fe/H]$_{phot}$ & \vgsr\ & dist$_{phot}$ \ & J$_{0}$\\
~& deg & deg & deg &deg& dex & \kms\ & kpc \ & mag\\
\hline
1&139.3878 & 29.1765 & 210.7694  &0.3501& -1.06 & 43.57 & 18.52 & 11.32 \\
2&202.0758 & -0.7620 & 277.4435  &6.4995 &-1.12 & 146.06 & 13.99 & 11.06\\
3&166.0221 & 20.5527 & 236.2520  &2.7682 &-0.94 & 85.43 & 8.08 & 10.19\\
4&190.1294 & 14.7341 & 259.5699  &1.3437 &-0.81 & 32.34 & 9.01 & 9.90\\
5&156.5275 & 17.3567 & 228.6722  &8.5960 &-1.07 & 30.51 & 10.23 & 10.25\\
6&174.4051 & 8.6444 & 248.2857  & 10.8177&-0.79 & 81.75 & 14.59 & 10.95\\
7&178.1447 & 15.4164 & 248.8794  & 3.1449&-0.97& 110.10 & 10.17 & 10.47\\
8&204.2222 & 13.1178 & 272.3394  &6.5864 &-0.91 & 178.71 & 15.35 & 11.42\\

\hline
\end{tabular}
\end{center}
\end{table}

\begin{thebibliography}{}
\bibitem[Belokurov et al.(2006)]{Bel06} Belokurov, V., Zucker, D.~B., Evans, N.~W., et al.\ 2006, \apjl, 642, L137 
\bibitem[Koposov \& Belokurov(2008)]{koposov2008} Koposov, S., \& Belokurov, V.\ 2008, Galaxies in the Local Volume, 195
\bibitem[Koposov et al.(2012)]{Koposov2012} Koposov, S.~E., Belokurov, V., Evans, N.~W., et al.\ 2012, \apj, 750, 80 
\bibitem[Belokurov et al.(2014)]{Belokurov2014} Belokurov, V., Koposov, S.~E., Evans, N.~W., et al.\ 2014, \mnras, 437, 116 
\bibitem[Bonaca et al.(2012)]{bonaca2012} Bonaca, A., Geha, M., \& Kallivayalil, N.\ 2012, \apjl, 760, L6
%\bibitem[Cooper et al.(2010)]{2010MNRAS.406..744C} Cooper, A.~P., Cole, S., Frenk, C.~S., et al.\ 2010, \mnras, 406, 744 
\bibitem[Chou et al.(2007)]{Chou2007} Chou, M.-Y., Majewski, S.R.,Cunha, K., et al.\ 2007, \apj, 670, 346
\bibitem[Chou et al.(2010)]{Chou2010} Chou, M.-Y., Cunha, K., Majewski, S.~R., et al.\ 2010, \apj, 708, 1290 
\bibitem[Cui et al.(2012)]{cui2012} Cui, X.-Q., Zhao, Y.-H., Chu, Y.-Q., et al.\ 2012, Research in Astronomy and Astrophysics, 12, 1197 
\bibitem[Davenport et al.(2014)]{Davenport2014} Davenport, J.~R.~A., Ivezi{\'c}, {\v Z}., Becker, A.~C., et al.\ 2014, \mnras, 440, 3430 
\bibitem[Deason et al.(2014)]{Deason2014} Deason, A.~J., Belokurov, V., Hamren, K.~M., et al.\ 2014, \mnras, 444, 3975 
\bibitem[Dehnen \& Binney(1998)]{1998MNRAS.298..387D} Dehnen, W., \& Binney, J.~J.\ 1998, \mnras, 298, 387 

\bibitem[Deng et al.(2012)]{2012RAA....12..735D} Deng, L.-C., Newberg, H.~J., Liu, C., et al.\ 2012, Research in Astronomy and Astrophysics, 12, 735 

\bibitem[Duffau et al.(2014)]{Duffau2014} Duffau, S., Vivas, A.~K., Zinn, R., M{\'e}ndez, R.~A., \& Ruiz, M.~T.\ 2014, \aap, 566, A118 

\bibitem[G{\'o}mez et al.(2013)]{gomez2013} G{\'o}mez, F.~A., Minchev, I., O'Shea, B.~W., et al.\ 2013, \mnras, 429, 159 

\bibitem[Grillmair et al.(2008)]{2008ApJ...689L.117G} Grillmair, C.~J., Carlin, J.~L., \& Majewski, S.~R.\ 2008, \apjl, 689, L117 
\bibitem[Helmi et al.(1999)]{helmi1999} Helmi, A., White, S.~D.~M., de Zeeuw, P.~T., \& Zhao, H.\ 1999, \nat, 402, 53

\bibitem[Helmi(2005)]{helmi2005} Helmi, A.\ 2005, The Identification of Dark Matter, 87 
\bibitem[Juri{\'c} et al.(2008)]{2008ApJ...673..864J} Juri{\'c}, M., Ivezi{\'c}, {\v Z}., Brooks, A., et al.\ 2008, \apj, 673, 864 
\bibitem[Klypin et al.(1999)]{klypin1999} Klypin, A., Kravtsov, A.~V., Valenzuela, O., \& Prada, F.\ 1999, \apj, 522, 82
\bibitem[Li et al.(2012)]{2012ApJ...757..151L} Li, J., Newberg, H.~J., Carlin, J.~L., et al.\ 2012, \apj, 757, 151
\bibitem[Li et al.(2016)]{2016arXiv160300262L} Li, J., Smith, M.~C., Zhong, J., et al.\ 2016, arXiv:1603.00262
\bibitem[Liu et al.(2014)]{liu14} Liu, C., Deng, L.-C., Carlin, J.~L., et al.\ 2014, \apj, 790, 110 
\bibitem[Law \& Majewski(2010)]{2010ApJ...714..229L} Law, D.~R., \& Majewski, S.~R.\ 2010, \apj, 714, 229 

\bibitem[Luo et al.(2015)]{Luo2015} Luo, A.-L., Zhao, Y.-H., Zhao, G., et al.\ 2015, Research in Astronomy and Astrophysics, 15, 1095 

\bibitem[Majewski et al.(2003)]{Majewski2003} Majewski, S.~R., Skrutskie, M.~F., Weinberg, M.~D., \& Ostheimer, J.~C.\ 2003, \apj, 599,1082
\bibitem[Majewski et al.(2004)]{Majewski2004} Majewski, S.~R., Kunkel, W.~E., Law, D.~R., et al.\ 2004, \aj, 128, 245
\bibitem[Mateo et al.(1996)]{Mateo1996} Mateo, M., Mirabal, N., Udalski, A., et al.\ 1996, \apjl, 458, L13 
\bibitem[Newberg et al.(2002)]{nyetal02} Newberg, H.~J., et al.\ 2002, \apj, 569, 245 
\bibitem[Newberg et al.(2007)]{heidi07} Newberg, H.~J., Yanny, B., Cole, N., et al.\ 2007, \apj, 668, 221 
\bibitem[Newberg et al.(2009)]{2009ApJ...700L..61N} Newberg, H.~J., Yanny, B., \& Willett, B.~A.\ 2009, \apjl, 700, L61 

\bibitem[Pila-D{\'{\i}}ez et al.(2014)]{Pila2014} Pila-D{\'{\i}}ez, B., Kuijken, K., de Jong, J.~T.~A., Hoekstra, H., \& van der Burg, R.~F.~J.\ 2014, \aap, 564, A18 

\bibitem[Pe{\~n}arrubia et al.(2010)]{Penarrubia2010} Pe{\~n}arrubia, J., Belokurov, V., Evans, N.~W., et al.\ 2010, \mnras, 408, L26
\bibitem[Rocha-Pinto et al.(2004)]{rochapinto2004} Rocha-Pinto, H.~J., Majewski, S.~R., Skrutskie, M.~F., Crane, J.~D., \& Patterson, R.~J.\ 2004, \apj, 615, 732 
\bibitem[Sanderson et al.(2012)]{sanderson2012} Sanderson, R.~E., Mohayaee, R., \& Silk, J.\ 2012, \mnras, 420, 2445 
\bibitem[Schlafly et al.(2011)]{Schlafly2011} Schlafly, E.~F., Finkbeiner, D.~P.\ 2011, \apj, 737, 103
\bibitem[Schlegel et al.(1998)]{Schlegel1998} Schlegel, D.~J., Finkbeiner, D.~P., \& Davis, M.\ 1998, \apj, 500, 525 
\bibitem[Sheffield et al.(2014)]{sheffield2014} Sheffield, A.~A., Johnston, K.~V., Majewski, S.~R., et al.\ 2014, \apj, 793, 62
\bibitem[Yanny et al.(2003)]{2003ApJ...588..824Y} Yanny, B., Newberg, H.~J., Grebel, E.~K., et al.\ 2003, \apj, 588, 824
\bibitem[Yanny et al.(2009)]{2009ApJ...700.1282Y} Yanny, B., Newberg, H.~J., Johnson, J.~A., et al.\ 2009, \apj, 700, 1282 

%\bibitem[Carlin et al.(2015)]{jeff15} Carlin, J.~L., Liu, C., Newberg, H.~J., et al.\ 2015, \aj, 150, 4 
textbf{\bibitem[Zhong et al.(2015)]{ZJ15} Zhong, J., L{\'e}pine, S., Li, J., et al.\ 2015, Research in Astronomy and Astrophysics, 15, 1154 }
\bibitem[Zhao et al.(2012)]{zhao2012} Zhao, G., Zhao, Y.-H., Chu, Y.-Q., Jing, Y.-P., \& Deng, L.-C.\ 2012, Research in Astronomy and Astrophysics, 12, 723 


\end{thebibliography}
\end{document}